\documentclass[%
 aip,
 amsmath,amssymb,
preprint,
]{revtex4-2}

\usepackage{graphicx}% Include figure files
\usepackage{dcolumn}% Align table columns on decimal point
\usepackage{bm}% bold math
\usepackage{hyperref}
\hypersetup{
    colorlinks = true,
    linkcolor = cyan,
    anchorcolor = cyan,
    citecolor = cyan,
    filecolor = cyan,
    urlcolor = cyan
    }
\usepackage{float}
\usepackage[utf8]{inputenc}
\usepackage[T1]{fontenc}
\usepackage{mathptmx}
\usepackage{etoolbox}
\usepackage{setspace}

\makeatletter
\def\@email#1#2{%
 \endgroup
 \patchcmd{\titleblock@produce}
  {\frontmatter@RRAPformat}
  {\frontmatter@RRAPformat{\produce@RRAP{*#1\href{mailto:#2}{#2}}}\frontmatter@RRAPformat}
  {}{}
}%
\makeatother
\begin{document}

\preprint{AIP/123-QED}

\title{A reduced-order mean-field synchronization model for thermoacoustic systems}
% Force line breaks with \\
\author{Rohan K. Nakade}
\affiliation{Department of Aerospace Engineering, Indian Institute of Technology Madras, Chennai 600036, India}
\affiliation{Centre of Excellence for Studying Critical Transition in Complex Systems, Indian Institute of Technology Madras, Chennai 600036, India}
 % \alffiliation[Also at ]{Physics Department, XYZ University.}%Lines break automatically or can be forced with \\
 
\author{Samarjeet Singh}%
\altaffiliation[Current address: ]{Engler-Bunte-Institute, Division of Combustion Technology, Karlsruhe Institute of Technology, Karlsruhe, 76131, Germany}
\affiliation{Department of Aerospace Engineering, Indian Institute of Technology Madras, Chennai 600036, India}
\affiliation{Centre of Excellence for Studying Critical Transition in Complex Systems, Indian Institute of Technology Madras, Chennai 600036, India}
 % \email{Second.Author@institution.edu.}
% \affiliation{ 
% Authors' institution and/or address%\\This line break forced with \textbackslash\textbackslash
% }%

\author{Jayesh M. Dhadphale}
\affiliation{Department of Aerospace Engineering, Indian Institute of Technology Madras, Chennai 600036, India}
\affiliation{Centre of Excellence for Studying Critical Transition in Complex Systems, Indian Institute of Technology Madras, Chennai 600036, India}
 % \homepage{http://www.Second.institution.edu/~Charlie.Author.}
% \affiliation{%
% Second institution and/or address%\\This line break forced% with \\
% }%
\author{R. I. Sujith}
\affiliation{Department of Aerospace Engineering, Indian Institute of Technology Madras, Chennai 600036, India}
\affiliation{Centre of Excellence for Studying Critical Transition in Complex Systems, Indian Institute of Technology Madras, Chennai 600036, India}

\date{\today}% It is always \today, today,

\begin{abstract}
% An article usually includes an abstract, a concise summary of the work covered at length in the main body of the article. It is used for secondary publications and for information retrieval purposes.
The synchronization phenomena in thermoacoustic systems leading to oscillatory instability can effectively be modeled using Kuramoto oscillators. Such models consider the nonlinear response of flame as an ensemble of Kuramoto phase oscillators constrained to collectively evolve at the rhythm of acoustic fluctuations. However, these high-dimensional models are analytically intractable and computationally expensive. In this study, we reduce the dimensionality of such a high-dimensional model and present a low-order, analytically tractable model for predicting transitions to thermoacoustic instability. We reduce the dimensionality of the phase oscillator model coupled to the acoustic field using the Ott-Antonsen ansatz [Chaos \textbf{18}, 037113 (2008)]. 
%Using the reduced-order equation, we compare and contrast the transition to thermoacoustic instability from the experiments. 
Using the reduced-order equations, we estimate the transitions to thermoacoustic instability and compare these transitions with the experiment. We validate the model for two combustor configurations, viz., the bluff-body stabilized dump combustor and the swirl-stabilized annular combustor. The low-order model accurately captures the continuous and abrupt secondary transitions observed experimentally in these distinct combustors.

\end{abstract}

\maketitle

\begin{quotation}
Combustion in confined spaces leads to the manifestation of oscillatory instabilities under specific conditions. These instabilities are often encountered as large-amplitude periodic pressure oscillations in combustors used for power generation in gas turbine and rocket engines. These instabilities are known as thermoacoustic instabilities in the combustion community and arise as a consequence of a positive feedback between the acoustic field in the combustor and the heat release rate of the flame. Such high amplitude oscillations may cause severe damage to the combustors. The acoustic pressure fluctuations are aperiodic and low-amplitude during the stable combustor operation. However, when a control parameter such as the Reynolds number or the equivalence ratio is varied, the system transitions to the state of thermoacoustic instability.
Depending on the operating parameters, this transition can be continuous or abrupt. 
To predict the onset of instability, a model for the heat release rate of the flame is required. Traditionally, the flame response to incoming perturbations is modeled using transfer functions obtained experimentally or numerically. Recently, synchronization theory has been used to study and model the synchronization phenomenon between the acoustic field and the heat release rate fluctuations from the reactive flow field. One such model based on the synchronization theory is the mean-field thermoacoustic model, which considers the flame as a population of Kuramoto phase oscillators coupled with each other and the acoustic field. The mean-field thermoacoustic model effectively captures the nature of transitions; however, the model is inherently high-dimensional and analytically intractable. We demonstrate a method to reduce the dimensionality of the mean-field model using the reduction method proposed by Ott and Antonsen [Chaos \textbf{18}, 037113 (2008)]. We show a way to approximate the frequency distribution obtained from the heat release rate spectrum with Lorentzian distributions and the capability of the reduced-order model to capture the nature of transitions.
\end{quotation}

% \doublespacing

\section{\label{introduction}Introduction}
Combustion in a confinement, such as a combustor, may lead to the occurrence of thermoacoustic instability, as a consequence of the positive feedback between the heat source and the acoustic field in the confinement. \cite{sujith_book, lieuwen2005combustion, culick2006unsteady} The occurrence of thermoacoustic instability leads to large amplitude pressure and velocity fluctuations. The unsteady flow in the combustor causes fluctuations in flame, generating sound waves that get reflected from the boundaries of the combustor. In turn, the reflected sound waves affect the flame, creating a positive feedback and amplifying the pressure and heat release rate oscillations. \cite{sujith_book} When heat is added during the compression or removed during the rarefaction phase of the pressure oscillations, respectively, energy is added to the acoustic field. \cite{Rayleigh1878} If this energy addition into the acoustic field exceeds the net acoustic losses, the acoustic oscillations get amplified. \cite{chu1965energy} These high-amplitude pressure oscillations have been a major issue in gas turbine and rocket engines, hindering their development and performance. Such high-pressure oscillations may cause severe structural damage to the combustor. The high amplitude oscillations lead to increased heat transfer, which overwhelms the thermal protection system. Due to the large amplitude vibrations, electronics, guidance and control systems may get damaged.

In a turbulent combustor, the stable combustor operation is characterized by aperiodic low-amplitude pressure fluctuations, called combustion noise in the combustion community. \cite{DOWLING201565} The time series of acoustic pressure fluctuations during combustion noise exhibits a broadband amplitude spectrum. On the other hand, the periodic pressure oscillations during thermoacoustic instability exhibit a narrowband amplitude spectrum. 

The system transitions from stable combustor operation to the state of thermoacoustic instability when a control parameter such as the Reynolds number ($Re$) or the equivalence ratio ($\phi$) is varied. \citet{lieuwen2002experimental} described this transition 
% from a state of stable operation to a state of unstable operation 
as a Hopf bifurcation from a stable, noisy fixed point to a distorted elliptic limit cycle, which can either happen in a continuous manner (supercritical bifurcation) or through an abrupt jump (subcritical bifurcation). Lieuwen observed these two transitions in the same system under distinct operating parameters. This description of the stable state (combustion noise) as a fixed point is valid for laminar systems. However, in turbulent combustors, \citet{nair2013loss} and \citet{tony2015detecting} showed that the stable state is high-dimensional chaos. 

Through a systematic variation of Reynolds number, \citet{nair2014intermittency} showed that the transition from combustion noise to thermoacoustic instability happens via a route of intermittency in both swirl-stabilized and bluff-body-stabilized configurations of backward-facing step combustors. During intermittency, the time series of acoustic pressure fluctuations exhibits bursts of periodic high-amplitude oscillations interspersed between epochs of low-amplitude aperiodic fluctuations. \citet{ananthkrishnan1998application} theoretically hypothesized about secondary bifurcation to large amplitude limit cycle oscillations. Later, such secondary bifurcations were observed in various thermoacoustic systems. \cite{roy2021flame, wang2021, samar_jegtp, bhavi2023abrupt} In such cases, the system transitions from combustion noise to low-amplitude thermoacoustic instability via intermittency and subsequently undergoes an abrupt jump to high-amplitude thermoacoustic instability.

Various models have been developed to study the transition of thermoacoustic systems from stable combustor operation to the state of thermoacoustic instability. One of the ways is to use computational fluid dynamics (CFD) to model all the relevant turbulent combustion processes. Nevertheless, CFD simulations are computationally expensive. Therefore, the flame transfer function (FTF) is used to model the response of the flame to the incoming perturbations. FTF establishes a link between the heat release rate fluctuations and the perturbations in velocity and equivalence ratio. \cite{FTF} However, FTF assumes the interactions to be linear and hence cannot describe the underlying nonlinear behavior of the system. The flame describing function (FDF) considers the nonlinear interactions by considering the amplitude and frequency of the incoming perturbations. \cite{FDF} 

Thermoacoustic systems often comprise acoustic interactions between various components such as ducts, nozzles, burners, flames, etc. These components can be approximated as a network of acoustic elements. \cite{polifke2010low, eckstein2006low, BOMBERG20153185} This approach of approximating the components as acoustic elements is called low-order network modeling and is a useful tool for analyzing the stability of thermoacoustic systems. In the low-order network models, the transfer matrix is constructed by using the coupling relations for the unknowns across each element. The system matrix is obtained by combining the transfer matrix coefficients. The system matrix is then used to obtain the eigenfrequencies and evaluate the stability characteristics of the system. Another way to study the transition from stable combustor operation to thermoacoustic instability is to model the flame dynamics using analytical models. The most widely used model is the $n-\tau$ model \cite{crocco1969research} which takes into account the time lag associated with combustion. However, the $n-\tau$ model does not account for the nonlinear nature of flame dynamics. Models such as the modified King's law \cite{heckl1990non} and the Levine-Baum model \cite{ananthkrishnan2005reduced} take into account the nonlinear relation between the heat release rate and the velocity fluctuations as well as the time lag associated with combustion. Once the heat release rate is modeled, the equations governing the flow field in the system can be set up and solved using various methods, such as the Galerkin modal expansion \cite{lores1973nonlinear} and the Green's function method. \cite{HECKL2007} 

\citet{dutta2019investigating} proposed a heat release rate model using phase oscillators, where the flame is considered as an ensemble of Kuramoto phase oscillators. In this model, the phase oscillators are coupled to the phase of the acoustic field, and the coupling strength depends on the normalized acoustic amplitude. This model is based on the mean-field synchronization of the phase oscillators and captures the various dynamical states (combustion noise, intermittency, and limit cycles) observed experimentally in a swirl-stabilized rotating swirler combustor. More recently, \citet{singh2024continuous} introduced the effect of acoustic coupling on the heat release rate, modeled using Kuramoto oscillators to develop a thermoacoustic model. This model explains the temporal and spatiotemporal aspects observed in distinct combustor configurations, viz.,  swirl-stabilized annular combustor, swirl-stabilized, and bluff-body-stabilized configurations of backward-facing step combustor. The natural frequencies of the oscillators were obtained from the Fast Fourier Transform (FFT) of the heat release rate oscillations during the dynamical state of combustion noise. Using parametric optimization, they found a linear relation between the equivalence ratio and the coupling strength. The model could accurately predict the various dynamical states observed experimentally for all the combustors. The mean-field model can predict not only the transition to thermoacoustic instability but also its suppression. This was demonstrated by \citet{samar2023openloop} where the model accurately predicted the suppression of thermoacoustic instability observed in experiments when the swirler rotation rate was systematically increased. 

%Through parametric optimization, the coupling strength was found to be linearly dependent on the swirler rotation rate.

The mean-field model of synchronization accurately predicts the nature of the transition. However, the number of oscillators considered for modeling the heat source is fairly high, making the model computationally expensive and analytically intractable. From \citet{singh2024continuous,samar2023openloop}, we observe that the nature of transition depends on the natural frequency distribution, or in other words, the FFT of heat release rate fluctuations during the occurrence of combustion noise. To predict the nature of transition, we propose an analytically tractable low-order model obtained by applying the Ott-Antonsen \cite{ott2008low} reduction method on the mean-field thermoacoustic model. We derive the low-order model by approximating the experimentally obtained natural frequency distribution using Lorentzian distributions. \cite{lorentzian} Using the low-order model, we estimate the nature of the transition, the bifurcation points and the normalized amplitudes. 

The subsequent paper is structured as follows: Section \ref{sec:experimental_setup} illustrates the experimental setup for two lab-scale turbulent combustors. Section \ref{sec:method} describes the mean-field thermoacoustic model, the Ott-Antonsen method for obtaining a low-order model, and a method for approximating the experimentally obtained natural frequency distribution using a mixture of Lorentzian distributions. We validate the low-order model using the \textit{N}-oscillator model in Sec. \ref{sec:validation}. Section \ref{sec:Results} shows the comparison of experimental and model results. Finally, Sec. \ref{sec:conclusion} summarizes the paper.

\section{\label{sec:experimental_setup}Experimental Setup}
\begin{figure*}
    \centering
    \includegraphics[width=\linewidth]{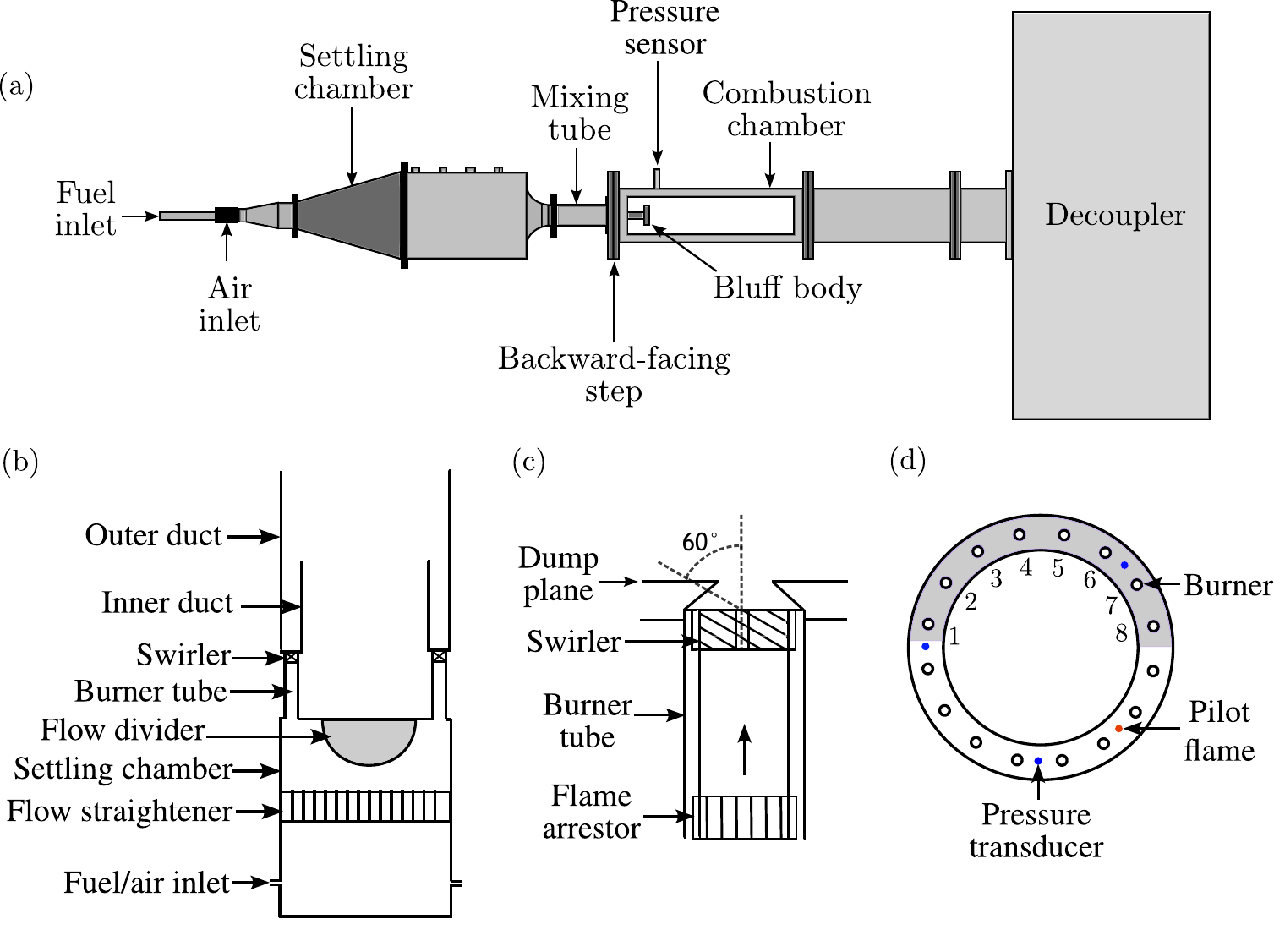}
    \caption{(a) Schematic of the bluff body stabilized dump combustor. Schematic of (b) the cross-section, (c) the burner tube and (d) the dump plane of the swirl stabilized annular combustor.}
    \label{fig:experimental setup}
\end{figure*}

\subsection{\label{sec:dump combustor}Bluff-body stabilized dump combustor}
The combustor comprises a settling chamber, a mixing duct, a combustion chamber, and a circular bluff body (see Figure \ref{fig:experimental setup}a). Air passes through the settling chamber, where the flow irregularities are removed, and enters the mixing tube. The fuel (liquified petroleum gas, 40\% propane and 60\% butane) enters the mixing tube via radial injection ports on the central shaft. Both air and fuel mix in the mixing tube and flow into the combustion chamber. 

The length of the combustion chamber is 1100 mm with a square cross-section of dimensions 90 $\times$ 90 mm$^2$. The combustion chamber near the bluff body has a backward-facing step. The reactants enter the combustion chamber from this side and are ignited with the help of a spark plug. The other end of the combustion chamber embodies a decoupler of dimensions 1000 $\times$ 500 $\times$ 500 mm$^3$, which isolates the combustion chamber from the ambient fluctuations outside. The flame is stabilized by the circular bluff body having a thickness of 10 mm and a diameter of 47 mm. The bluff body is located 32 mm downstream of the backward-facing step. A fixed fuel flow rate of 48 SLPM is maintained, with the airflow rate varying from 752 SLPM to 1446 SLPM. This leads to a variation of the Reynolds number $Re$ from 3.9 $\times$ $10^{4}$ to 6.4 $\times$ $10^{4}$ and the equivalence ratio $\phi$ from 1 to 0.52. Here, the Reynolds number is given as $Re = V d/ \nu$, where $V$ is the mean velocity of the flow, $d$ is the bluff-body diameter and $\nu$ is the kinematic viscosity. The equivalence ratio is defined as the ratio of the actual fuel-air ratio to the stoichiometric fuel-air ratio.

The heat release rate fluctuations are measured using the fluctuations in chemiluminescence intensity. \cite{hardalupas2004188, guethe2012chemiluminescence} These intensity fluctuations are acquired using a high-speed camera equipped with a CH$^{*}$ filter and a 100 mm Carl-Zeiss lens. The images were acquired at a sampling rate of 2 kHz at a resolution of 520 $\times$ 520 pixels. The noise effects were removed by coarse-graining the images by combining 10 $\times$ 10 pixels. A detailed description of the dump combustor setup and the measurements can be found in \citet{sudarsanan2024emergence}.

\subsection{\label{annular combustor}Swirl-stabilized annular combustor}
Figure \ref{fig:experimental setup}(b) shows a schematic of the annular combustor. The annular combustor comprises 16 burner tubes, each 30 mm in diameter and 150 mm long. A technically premixed fuel-air mixture enters the settling chamber through twelve equally spaced inlet ports. The fuel used was liquified petroleum gas. A honeycomb mesh in the settling chamber removes the non-uniformities in the flow. The flow is then guided uniformly in the 16 burner tubes through a hemispherical divider. Each burner tube has a swirler on the top for flame stabilization and a flame arrestor at the bottom for preventing flashback. Swirlers mounted on each burner comprise a central shaft (15 mm diameter) and six guide vanes, each making a 60$^{\circ}$ angle with the shaft axis (see Fig. \ref{fig:experimental setup}c).

The combustion chamber consists of two concentric ducts with inner and outer diameters of 300 mm and 400 mm, respectively. The length of the outer duct is 400 mm, and the inner duct is 200 mm. A non-premixed pilot flame (see Fig. \ref{fig:experimental setup}d) ignites the main combustor. The pilot flame is later extinguished after flame stabilization. The airflow rate is fixed at 1400 SLPM. The fuel flow rate is increased from 40 to 48 SLPM ($\phi$ = 0.82 to 0.98) in the forward direction and decreased from 48 to 36 SLPM ($\phi$ = 0.98 to 0.73) in the reverse direction. Considering the fixed airflow rate, which is much larger than the fuel flow rate, the Reynolds number is calculated to be $Re_d \approx 8600$ using the exit diameter of the burner tube ($d = 15$ mm).

The CH$^{*}$ chemiluminescence intensity fluctuations of swirling flames were captured using a high-speed camera. These intensity fluctuations are used to obtain the heat release rate oscillations. Images were acquired at a resolution of 1280 $\times$ 800 pixels at a sampling rate of 2 kHz. Only one-half of the combustor backplane (the shaded portion in Fig. \ref{fig:experimental setup}d) was imaged with the help of an air-cooled mirror 1 m above the combustor. The camera lens used was Nikkon AF Nikkor with 70 - 210 mm $f$/4 to $f$/5.6. Additional information about the annular combustor setup can be found in \citet{roy2021flame}, and \citet{singh2024intermittency}.

\subsection{Instrumentation}
The mass flow controllers (MFCs) used to regulate the flow rates of fuel and air in both the annular and dump combustor setups are Alicat Scientific MCR Series. These MFCs have an uncertainty of $\pm 0.8\%$ of the measured reading and $\pm 0.2\%$ of the full-scale reading. The maximum uncertainties in $Re$ and $\phi$ are $\pm 0.8 \%$ and $ \pm 1.6 \%$, respectively. A piezoelectric transducer PCB103B02, having a sensitivity of 217.5 mV/kPa and an uncertainty of $\pm 0.15$ Pa was used to obtain the pressure fluctuations. The signal is recorded for a duration of 3 s at a sampling frequency of 10 kHz for both combustor configurations. The chemiluminescence images were obtained using the high-speed camera (CMOS Phantom V12.1) fitted with a CH$^{*}$ filter. The fluctuations in global heat release rate were acquired using a photomultiplier tube (PMT) equipped with a CH$^{*}$ filter. The CH$^{*}$ filter used in both the camera and the PMT has a bandwidth of 435 $\pm$ 10 nm.

\section{\label{sec:method}Approach and methods}
\subsection{\label{sec:governing eqns}Governing equations for the acoustic field and the heat release rate}
The linearized momentum and energy equations in a one-dimensional thermoacoustic system with the negligible effect of temperature gradient and mean flow, driven by an oscillatory heat release rate, are \cite{balasubramanian_and_sujith,nicoud2009zero}
\begin{gather}
    \frac{1}{\tilde{\rho}_{0}} \frac{\partial \tilde{p}'}{\partial \tilde{z}} + \frac{\partial \Tilde{u}'}{\partial \Tilde{t}} = 0
    \label{momentum}, \\
    \frac{\partial \Tilde{p}'}{\partial \tilde{t}} + \gamma \tilde{p}_{0} \frac{\partial \Tilde{u}'}{\partial \Tilde{z}} = (\gamma - 1) \dot{\Tilde{q}}' \delta(\tilde{z} - \tilde{z}_{f}),
    \label{energy}
\end{gather}
where $\tilde{p}'$ and $\tilde{u}'$ are the mean subtracted fluctuations of the acoustic pressure and velocity, respectively. $\tilde{p}_{0}$ and $\tilde{\rho}_{0}$ are the pressure and density of the mean flow, $\tilde{z}$ denotes the axial distance along the duct, $\gamma$ is the ratio of specific heat capacities, and $\tilde{t}$ is time. $\dot{\tilde{q}}'$ represents the unsteady heat release rate fluctuations per unit area. These fluctuations are assumed to be acoustically compact with concentration at $\tilde{z} = \tilde{z}_{f}$, represented by the Dirac delta function $\delta(\tilde{z} - \tilde{z}_{f})$. $\tilde{()}$ indicates that the terms are dimensional.

The above partial differential equations can be transformed into ordinary differential equations by employing the Galerkin technique \cite{lores1973nonlinear}. In this method, the acoustic pressure and velocity fluctuations are expanded as a superposition of spatial basis functions (sine and cosine) with time-varying coefficients ($\eta(\tilde{t})$ and $\dot{\eta}(\tilde{t})$) that satisfy the given boundary conditions. We choose the basis functions satisfying the boundary conditions of an open-closed duct.
\begin{gather}
        \Tilde{p}'(\Tilde{z},\Tilde{t}) = \tilde{p}_{0}\sum_{j=1}^{n} \frac{\dot{\eta}_j(\Tilde{t})}{\Tilde{\Omega}_{j}} \cos(\Tilde{k}_{j}\Tilde{z})
        \label{pprime},\\
        \Tilde{u}'(\Tilde{z},\Tilde{t}) = \frac{\tilde{p}_{0}}{\tilde{c}_{0}\tilde{\rho}_{0}} \sum_{j=1}^{n} \eta_{j}(\Tilde{t}) \sin(\Tilde{k}_{j}\Tilde{z}).
        \label{uprime}
\end{gather}
Here $\tilde{c}_{0}$ is the average sound velocity in the duct, $\tilde{k}_{j} = (2j-1)\pi / (2\Tilde{L})$ is the wavenumber, $\Tilde{\Omega}_{j} = \tilde{c}_{0}\Tilde{k}_{j}$ represents the natural frequency for the $j^{th}$ mode of the system, and $\tilde{L}$ is the duct length. Substituting the expansions for $\tilde{p}'$ and $\tilde{u}'$ in Eq. (\ref{energy}), we get
\begin{equation}
     \sum_{j=1}^{n} \frac{\ddot{\eta}_j(\Tilde{t})}{\Tilde{\Omega}_{j}} \cos(\Tilde{k}_{j}\Tilde{z}) + \frac{\gamma \tilde{p}_{0}}{\tilde{c}_{0}\tilde{\rho}_{0}} \sum_{j=1}^{n} \tilde{k}_j \eta_{j}(\Tilde{t})  \cos(\Tilde{k}_{j}\Tilde{z}) = \frac{(\gamma - 1)}{\tilde{p}_{0}} \dot{\Tilde{q}}' \delta(\tilde{z} - \tilde{z}_{f}).
     \label{galerkin_imterim}
\end{equation}
The resulting equation is projected onto the basis functions by computing an inner product of Eq. (\ref{galerkin_imterim}) with $\cos(\tilde{k}_{j}\tilde{z})$. This involves multiplying both sides of Eq. (\ref{galerkin_imterim}) with $\cos(\tilde{k}_{j}\tilde{z})$ and integrating over the spatial domain from $0$ to $\tilde{L}$. As a result, the second-order ordinary differential equations for individual modes are given as
\begin{equation}
    \frac{\ddot{\eta}_{j}(\Tilde{t})}{\Tilde{\Omega}_{j}} + \tilde{c}_{0}\Tilde{k}_{j} \eta_{j}(\Tilde{t}) = \frac{2(\gamma-1)}{\Tilde{L} \tilde{p}_{0}} \int_{0}^{\Tilde{L}} \dot{\Tilde{q}}' \delta(\Tilde{z} - \Tilde{z}_{f}) cos(\Tilde{k}_{j}\Tilde{z})d\Tilde{z}.
\end{equation}
Here, we use $\tilde{c}_{0} = \sqrt{\gamma \tilde{p}_0/ \tilde{\rho}_0}$ and the identity, $\int_{0}^{\tilde{L}} cos^{2} (\Tilde{k}_{j}\Tilde{z}) d\Tilde{z} = \Tilde{L}/2$. We neglect the effects of higher modes and assume that the single-mode analysis captures the transitions reasonably well, \cite{Subramanian_Sujith_Wahi_2013} to obtain
\begin{equation}
    \Ddot{\eta}(\Tilde{t}) + \Tilde{\alpha}\Dot{\eta}(\Tilde{t}) +  \Tilde{\Omega}_{0}^{2}\eta(\Tilde{t}) = \frac{2(\gamma-1)}{\Tilde{L} \tilde{p}_{0}} \tilde{\Omega}_0 \int_{0}^{\Tilde{L}} \dot{\Tilde{q}}' \delta(\Tilde{z} - \Tilde{z}_{f}) \cos{(\Tilde{k}\Tilde{z})}d\Tilde{z},
    \label{single mode equation}
\end{equation}
where we introduce the term $\Tilde{\alpha}\Dot{\eta}(\Tilde{t})$ for accounting the acoustic damping of the system \cite{mateev_and_culick} with $\tilde{\alpha}$ being the damping coefficient.

A model for the heat release rate $\dot{\tilde{q}}'$ is required to obtain the acoustic field in the system. The mean field thermoacoustic model considers the flame response as a combined effect of a population of Kuramoto phase oscillators coupled with the acoustic field. \cite{dutta2019investigating, singh2024continuous} The dynamical equation for the $j^{th}$ phase  oscillator having a phase ($\theta_{j}$) at time $\tilde{t}$ and having a natural frequency ($\tilde{\omega}_{j}$) is
\begin{equation}
    \Dot{\theta}_{j}(\tilde{t}) = \tilde{\omega}_{j} + \tilde{K}\hat{R}(\tilde{t})\sin(\Phi(\tilde{t}) - \theta_{j}(\tilde{t})),
    \label{theta_dot_dim}
\end{equation}
where $\hat{R}$ is the normalized amplitude of the acoustic variable ($\eta$) and $\Phi$ is its phase. In this model, the term $\tilde{K}\hat{R}$ represents the effective coupling strength of the oscillators with the acoustic field, which signifies that high-amplitude acoustic fluctuations enhance coupling. The individual contribution of each Kuramoto phase oscillator is summed to obtain $\dot{\tilde{q}}'$ as
\begin{equation}
    \Dot{\Tilde{q}}' = \tilde{q}_{0} \sum_{j=1}^{N} \sin \big[ \Tilde{\Omega}_{0}\Tilde{t} + \theta_{j}(\Tilde{t}) \big],
    \label{heat release}
\end{equation}
where $\tilde{q}_{0}$ represents the heat release rate amplitude of individual oscillators and \textit{N} denotes the number of oscillators. Equation \eqref{heat release} indicates that the heat release rate fluctuations are high (low) when the oscillators are synchronized (unsynchronized). The level of synchrony is quantified by the order parameter $z$, defined as
\begin{equation}
    z = r e^{i\psi} = \dfrac{1}{N}\sum_{j=1}^{N}e^{i\theta_{j}} \,,
    \label{order parameter kuramoto}
\end{equation}
where $\psi$ and $r$ are the phase and magnitude of the complex order parameter $z$. Substituting the heat release rate from Eq. (\ref{heat release}) in Eq. (\ref{single mode equation}), the equation governing the acoustic field can be written as
\begin{equation}
    \Ddot{\eta}(\Tilde{t}) + \Tilde{\alpha}\Dot{\eta}(\Tilde{t}) +  \Tilde{\Omega}_{0}^{2}\eta(\Tilde{t}) = \tilde{\beta} \cos(\tilde{k} \tilde{z}_{f}) \sum_{j=1}^{N} \sin \big[ \tilde{\Omega}_{0}\tilde{t} + \theta_{j}(\tilde{t}) \big], \label{thermoacoustic_dim}
\end{equation}
where $\tilde{\beta} = 2\tilde{q}_{0}(\gamma - 1)/\tilde{L} \tilde{p}_{0}$ is the measure of flame strength. Equations (\ref{theta_dot_dim}) and (\ref{thermoacoustic_dim}) are non-dimensionalized using
\begin{equation}
    t = \tilde{\Omega}_0 \tilde{t}, \;\; \alpha = \dfrac{\tilde{\alpha}}{\tilde{\Omega}_{0}}, \;\; \beta = \dfrac{\tilde{\beta}}{\tilde{\Omega}_{0}}, \;\; \omega_{i} = \dfrac{\tilde{\omega}_{i}}{\tilde{\Omega}_{0}},  \;\; k = \tilde{k}\tilde{L}, \;\; z = \dfrac{\tilde{z}}{\tilde{L}}, \;\;  K = \dfrac{\tilde{K}}{\tilde{\Omega}_{0}}. \label{nondimensionalization}
\end{equation}
This leads to the following non-dimensionalized equations:
\begin{gather}
    \Dot{\theta}_{j}(t) = \omega_{j} + K\hat{R}(t)\sin(\Phi(t) - \theta_{j}(t))
    \label{theta_dot}, \\
    \Ddot{\eta}(t) + \alpha\Dot{\eta}(t) + \eta(t) = \beta \cos(k z_{f}) \sum_{j=1}^{N} \sin \big[t + \theta_{j}(t) \big] . \label{thermoacoustic}
\end{gather}

Assuming the fluctuations to be quasi-harmonic, we decompose the acoustic variable $\eta(t)$ as \cite{krylov1949introduction}
\begin{equation}
    \eta(t) = -R(t)\cos(t + \Phi(t)),
    \label{Krylov Bogolyubov decomposition}
\end{equation}
where $\Phi (t)$ and $R(t)$ are the phase and the envelope amplitude, which evolve slower as compared to the fast time scale $2\pi/\Omega_0$. Using this decomposition in Eq. (\ref{thermoacoustic}) yields a second-order ordinary differential equation in $R$ and $\Phi$. Further, the sines and cosines can be expressed in complex form using Euler's representation, and the fast time scales can be eliminated using the method of averaging \cite{balanov2009simple}. Equating the imaginary and real parts of the resulting differential equation leads to (see Appendix \ref{appendix: derivation of Rddot and phiddot} for detailed derivation)
\begin{gather}
    \Ddot{R} = 2R\dot{\Phi} + R\dot{\Phi}^{2} - \alpha \dot{R} - \beta \cos(kz_{f})\sum_{j=1}^{N}\sin(\theta_{j} - \Phi), \label{R_ddot_der}\\
    R\ddot{\Phi} = \beta \cos(kz_{f}) \sum_{j=1}^{N}\cos(\theta_{j} - \Phi) - 2\dot{R}\dot{\Phi} - \alpha R \dot{\Phi} - 2\dot{R} - \alpha R. \label{phi_ddot_der}
\end{gather}
By multiplying both sides of Eq. \eqref{order parameter kuramoto} by $e^{-i\Phi}$ and equating the imaginary and real parts of the resulting equation, we can express the summation of sines and cosines in terms of $r$ and $\psi$ as 
\begin{gather}
    \dfrac{1}{N}\sum_{j=1}^{N} \sin(\theta_j - \Phi) = r \, \sin(\psi-\Phi) \label{sin sum} \\
    \dfrac{1}{N}\sum_{j=1}^{N} \cos(\theta_j - \Phi) = r \, \cos(\psi-\Phi) \label{cos sum}
\end{gather}
Using Eqs. \eqref{sin sum} and \eqref{cos sum} in Eqs. \eqref{R_ddot_der} and \eqref{phi_ddot_der} leads to
\begin{gather}
    \Ddot{R} = 2R\dot{\Phi} + R\dot{\Phi}^{2} - \alpha \dot{R} - \beta \,N\, \cos(kz_{f})\,r\,\sin(\psi - \Phi), \label{R_ddot_nonnormalized}\\
    R\ddot{\Phi} = \beta \,N\, \cos(kz_{f})\, r \,\cos(\psi - \Phi) - 2\dot{R}\dot{\Phi} - \alpha R \dot{\Phi} - 2\dot{R} - \alpha R. \label{phi_ddot_nonnormalized}
\end{gather}

We now normalize the envelope amplitude ($R$) using the amplitude of limit cycle oscillations ($R_{LCO}$) at $K \to \infty$. To compute $R_{LCO}$, we transform Eqs. \eqref{R_ddot_nonnormalized} and \eqref{phi_ddot_nonnormalized} to Cartesian coordinates using the following transformation: 
\begin{gather}
    A = R\cos(\Phi), \;\;\; B = R\sin(\Phi). \label{cartesian transform}
\end{gather}
The transformed equations become
\begin{gather}
    \ddot{A} = - \alpha \dot{A} + 2\dot{B} + \alpha B -\beta \,N\, \cos(kz_{f})\, r\sin(\psi) \label{A_ddot}, \\
    \ddot{B} = -2\dot{A} -\alpha \dot{B} - \alpha A + \beta \,N\, \cos(kz_{f}) \,r \cos(\psi). \label{B_ddot}
\end{gather}
The limit cycle solution is given by $\ddot{A}=\dot{A}=0$ and $\ddot{B}=\dot{B}=0$. For $K \to \infty$, the oscillators will be synchronized, and the magnitude of the order parameter will be $r = 1$. Using these conditions in Eqs. (\ref{A_ddot}) and (\ref{B_ddot}) leads to the following: 
\begin{gather}
    A_{LCO} = \dfrac{\beta \,N\, \cos(kz_{f})\, \cos(\psi)}{\alpha}, \label{A}\\
    B_{LCO} = \dfrac{\beta \,N\, \cos(kz_{f})\, \sin(\psi)}{\alpha}. \label{B}
\end{gather}
By using Eq. (\ref{cartesian transform}) and expressing $A$ and $B$ from Eqs. (\ref{A}) and (\ref{B}) in terms of $R$, we obtain,
\begin{gather}
    R_{LCO} = \dfrac{\beta \,N\, \cos(kz_{f})}{\alpha}.
\end{gather}
The acoustic amplitude is normalized as $\hat{R} = {R}/{R_{LCO}}$ in Eqs. (\ref{R_ddot_nonnormalized}) and (\ref{phi_ddot_nonnormalized}), resulting in the following equations,
\begin{gather}
    \Ddot{\hat{R}} = 2\hat{R}\dot{\Phi} + \hat{R}\dot{\Phi}^{2} - \alpha \dot{\hat{R}} - \alpha \, r\,\sin(\psi - \Phi), \label{R_ddot_normalized}\\
    \hat{R}\ddot{\Phi} = \alpha \, r \,\cos(\psi - \Phi) - 2\dot{\hat{R}}\dot{\Phi} - \alpha \hat{R} \dot{\Phi} - 2\dot{\hat{R}} - \alpha \hat{R}. \label{phi_ddot_normalized}
\end{gather}
The system of equations \eqref{theta_dot}, \eqref{R_ddot_normalized} and \eqref{phi_ddot_normalized} represent the mean-field thermoacoustic model and can be integrated in time to obtain the bifurcation diagrams.

\subsection{\label{sec:ott-antonsen}Ott-Antonsen reduction}
\begin{figure}
    \centering
    \includegraphics[width= 0.9\linewidth]{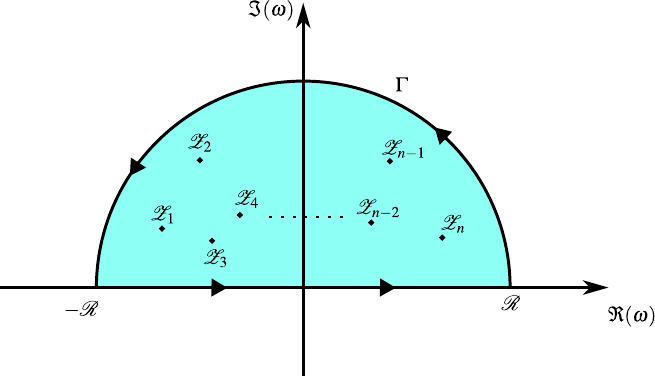}
    \caption{\doublespacing Schematic of the closed contour in complex $\omega$-plane considered for performing integration. The contour includes the line on real $\omega$ axis from $-\mathcal{R}$ to $\mathcal{R}$ ($\mathcal{R} \to \infty$) and the semicircle ($\Gamma$). The arrows on the contour indicate the direction of the integration path.}
    \label{fig:contour}
\end{figure}

The equations  \eqref{theta_dot}, \eqref{R_ddot_normalized} and \eqref{phi_ddot_normalized} have a dimensionality of $N$, 2 and 2, respectively, resulting in an overall dimensionality of \textit{N}+4, where \textit{N} is the number of oscillators. For a large number of oscillators, the model becomes analytically intractable and the simulations become computationally expensive. In this section, we utilize the Ott-Antonsen method to derive a reduced-order model.
To reduce the dimensionality of the system, a continuum limit of $N \rightarrow$ $\infty$ is considered, and a probability density function ($f$) is defined such that at any time $t$, $f(\theta,\omega,t)d\theta d\omega$ represents the fraction of oscillators having natural frequency between $\omega$ and $\omega + d\omega$ and phase between $\theta$ and $\theta + d\theta$. The probability density function satisfies the normalization,
\begin{equation}
\int_{-\infty}^{\infty} \int_{0}^{2\pi} f(\theta,\omega,t) d\theta d\omega = 1,
\end{equation}
which leads to,
\begin{equation}
    \int_{0}^{2 \pi} f(\theta, \omega, t) d\theta = g(\omega).
\end{equation}
Here, $g(\omega)$ is the time-independent natural frequency distribution of the oscillators. Since the number of oscillators is conserved, $f$ satisfies the continuity equation, \cite{ott2008low}
\begin{equation}
    \frac{\partial f}{\partial t} + \frac{\partial (fv)}{\partial \theta} = 0,
    \label{continuity}
\end{equation}
where $v = \dot{\theta}$ is the angular velocity of the oscillators given by Eq. \eqref{theta_dot}. The probability density function $f(\theta,\omega,t)$ can be expressed as a Fourier series expansion in $\theta$ as 
\begin{equation}
    f(\theta,\omega,t) = \frac{g(\omega)}{2\pi}[1 + \sum_{n=1}^{\infty} (a(\omega,t)^{n}e^{in\theta} + \overline{a}(\omega,t)^{n}e^{-in\theta})],
    \label{f}
 \end{equation}
where $a(\omega,t)$ is the Fourier series coefficient and $\overline{a}(\omega,t)$ is its complex conjugate. We assume $|a(\omega,t)|<1$ for ensuring convergence. This form of the probability density function, with the Fourier coefficients as a power series of $a(\omega,t)$, naturally connects the incoherent and partially synchronized states. \cite{ott2008low} Substituting $f$ and $v$ from Eqs. \eqref{f} and \eqref{theta_dot} in Eq. \eqref{continuity} gives
\begin{equation}
    \frac{\partial \overline{a}}{\partial t} -i\omega \overline{a} - \frac{K\hat{R}}{2}e^{i\Phi} + \frac{K\hat{R}}{2}e^{-i\Phi}\overline{a}^{2} = 0.
    \label{diffa}
\end{equation}
In the continuum limit ($N \to \infty$), the order parameter defined in Eq. \eqref{order parameter kuramoto} is expressed as
\begin{equation}
    z = re^{i\psi} = \int_{-\infty}^{\infty} \int_{0}^{2\pi} f(\theta,\omega,t) e^{i\theta} d\theta d\omega .
    \label{orderparam}
\end{equation} 
Substituting the Fourier expansion of $f$ (Eq. \ref{f}) in the above equation leads to
\begin{equation}
    z = re^{i\psi} = \int_{-\infty}^{\infty} \overline{a}(\omega,t) g(\omega) d\omega.
    \label{order param integral}
\end{equation}
The natural frequency distribution $g(\omega)$, which represents the density of oscillators with a natural frequency $\omega$, must be known to compute the above integral. \citet{singh2024continuous} utilized the FFT of chemiluminescence data during the combustion noise state as the natural frequency distribution of the oscillators. This distribution can be effectively approximated using a weighted sum of Lorentzian distributions \cite{sowmiya2020cauchy} as
\begin{equation}
    g(\omega) = \sum_{k=1}^{n} \dfrac{c_{k}}{\pi}\dfrac{\gamma_{k}}{\left[ (\omega - \omega_{k})^{2} + \gamma_{k}^{2}\right]},
    \label{lorentz_mix}
\end{equation}
by minimizing Kullback-Leibler (KL) divergence. Here, $\omega_{k}$ is the peak location, $\gamma_{k}$ is the half-width at half-maximum and $c_{k}$ is the weight for the $k^{th}$ Lorentzian distribution. $n$ denotes the total number of Lorentzian distributions used to approximate the experimental distribution. The natural frequency distribution satisfies the normalization, $\int_{-\infty}^{\infty} g(\omega) d\omega = 1$ which leads to
\begin{equation}
    \sum_{k=1}^{n}c_{k} = 1.
    \label{fd normalization}
\end{equation}
Lorentzian distributions are chosen for the fitting because they contain simple poles, which makes the computation of integrals straightforward. Using the frequency distribution from Eq. (\ref{lorentz_mix}), the integration in Eq. (\ref{order param integral}) is computed with the aid of the residue theorem. A closed contour (Fig. \ref{fig:contour}) is selected, consisting of the real $\omega$-axis and a semi-circle with a large radius ($\mathcal{R} \to \infty$) in the upper half of the complex $\omega$-plane ($\Im(\omega) > 0$). The poles of the frequency distribution inside this contour are given by $\mathcal{Z}_{k} = \omega_{k} + i\gamma_{k}$. Using these poles in the residue theorem, the integration in Eq. \eqref{order param integral} simplifies to
\begin{equation}
    z = \sum_{k=1}^{n}c_{k} \; \overline{a}(\omega_{k} + i\gamma_{k},t).
    \label{z_aft_int}
\end{equation}
The complex Fourier coefficients $\overline{a}(\omega_{k}+i\gamma_{k},t)$ are expressed in polar forms as $r_{k}\exp(i\phi_{k})$ and substituted in Eq. \eqref{diffa}. By comparing the imaginary and real parts on both sides of the resulting equation, we obtain the following set of equations
\begin{gather}
    \Dot{r}_{k} = -r_{k}\gamma_{k} + \dfrac{K\hat{R}}{2}(1 - r_{k}^{2})\cos(\phi_{0k}) \label{rk_dot},\\
    \Dot{\phi}_{k} = \omega_{k} - \dfrac{K\hat{R}}{2r_{k}}(1 + r_{k}^{2})\sin(\phi_{0k}), \label{phi0k_dot}
\end{gather}
where,
\begin{gather}
    \phi_{0k} = \phi_{k} - \Phi.
    \label{phi_0k equation}
\end{gather}
The fixed-point solutions of the system are obtained by equating the time derivatives to zero. For the system of Eqs. (\ref{R_ddot_normalized}), (\ref{phi_ddot_normalized}), and (\ref{rk_dot}) - (\ref{phi_0k equation}), the fixed point corresponds to the state where $\dot{r}_k$, $\dot{\phi}_{0k}$, $\dot{\hat{R}}$, $\ddot{\hat{R}}$, and $\ddot{\Phi}$ are zero. These equations are nonlinear and require numerical methods to solve them. We use the arclength continuation method, \cite{sieber2016ddebiftoolmanualbifurcation, ddebiftool} where the solution curve is traced iteratively. The initial guess is computed by taking a small step along the tangent of the curve for each iteration. This guess is then updated along a circle with a radius equal to the step length until convergence. Once the $r_{k}$ and $\phi_{k}$ at the fixed point solutions are computed, the order parameter can be obtained using Eq. \eqref{z_aft_int}. One key advantage of the arclength continuation method over the commonly used Newton-Raphson method is its ability to obtain all solutions, even when the function exhibits folds \cite{keller1987lectures}. This enables us to capture the unstable limit cycle solutions.

\subsection{\label{sec:trivial soln}The trivial solution}
The system has a trivial solution $r_n = \hat{R} = 0$, which cannot be numerically computed using the arclength continuation method due to ill-conditioning of the Jacobian matrix. To obtain the stability characteristics at these trivial fixed points, we transform the system to Cartesian coordinates using $\overline{a}(\omega_{k}+i\gamma_{k},t) = x_{k} + iy_{k}$ in Eq. (\ref{diffa}). This transformation leads to
\begin{gather}
    \dot{x}_{k} = -\gamma_{k}x_{k} - \omega_{k}y_{k} + \dfrac{K \hat{A}}{2} - \dfrac{K}{2} \left[ \hat{A}(x_{k}^{2} - y_{k}^{2}) + 2 \hat{B}x_{k}y_{k} \right] \label{xk_dot},\\
    \dot{y}_{k} = \omega_{k}x_{k} - \gamma_{k}y_{k} + \dfrac{K\hat{B}}{2} - \dfrac{K}{2} \left[ 2\hat{A}x_{k}y_{k} - \hat{B}(x_{k}^{2} - y_{k}^{2}) \right]. \label{yk_dot}
\end{gather}
Equations (\ref{cartesian transform}), (\ref{A_ddot}) and (\ref{B_ddot}) can be normalized using the limit cycle amplitude ($R_{LCO}$) as
\begin{gather}
    \hat{A} = \hat{R}\cos(\Phi), \;\;\;\hat{B} = \hat{R}\sin(\Phi) \label{cartesian_normalized},\\
    \ddot{\hat{A}} = -\alpha \dot{\hat{A}} + 2\dot{\hat{B}} + \alpha \hat{B} - \alpha r \sin(\psi) \label{A_ddot_interim},\\
    \ddot{\hat{B}} = - 2\dot{\hat{A}} - \alpha \dot{\hat{B}}  + \alpha \hat{A} + \alpha r \cos(\psi) \label{B_ddot_interim}.
\end{gather}
From Eq. \eqref{z_aft_int}, we have
\begin{gather}
    z = re^{i\psi} = r \cos(\psi) + i \, r \sin(\psi) = \sum_{k=1}^{n}c_{k}(x_{k} + i y_{k}). \label{z_in_xy}
\end{gather}
Using Eqs. \eqref{A_ddot_interim}, \eqref{B_ddot_interim} and \eqref{z_in_xy}, we obtain
\begin{gather}
    \ddot{\hat{A}} = -\alpha \dot{\hat{A}} + 2\dot{\hat{B}} + \alpha \hat{B} - \alpha \sum_{k=1}^{n}c_{k}y_{k} \label{A_ddot_normalized},\\
    \ddot{\hat{B}} = - 2\dot{\hat{A}} - \alpha \dot{\hat{B}}  + \alpha \hat{A} + \alpha \sum_{k=1}^{n}c_{k}x_{k}  \label{B_ddot_normalized}.
\end{gather}
Equations (\ref{xk_dot}), (\ref{yk_dot}), (\ref{A_ddot_normalized}) and (\ref{B_ddot_normalized}) have a trivial fixed-point solution at $x_k = y_k = \hat{A} = \hat{B} = 0$, where the time-derivatives become zero. The stability of these fixed points can be evaluated by calculating the eigenvalues of the Jacobian at these points. 

\subsection{\label{sec:bifurcation points}Bifurcation points}
The stability of a fixed-point solution is determined by computing the eigenvalues of the Jacobian at that point. For a stable fixed point, all eigenvalues must have a negative real part. As we move along the solution branch, the eigenvalues vary smoothly, and some will cross the imaginary axis when the stability of the system changes. Therefore, the locations on the solution branch where the real part of any eigenvalue becomes zero correspond to the locations of bifurcation points \cite{strogatz2018nonlinear}.

Consider a dynamical system $\dot{X} = f(X)$, where $X = [x_{1}, x_{2}, ..., x_{n}]^T$ is the state of the system at any time $t$ and $f = [f_{1}, f_{2}, ..., f_{n}]^T$ is a differentiable function. The fixed-point solution $X^{*}$ is given as $f(X^{*}) = 0$. The Jacobian $J(X)$ of $f(X)$ and its eigenvalues $\lambda (X)$ at any state $X$ are given by
\begin{gather}
    J(X) = \dfrac{d}{dX}f(X), \label{Jacobian}\\
    |J(X) - \lambda(X) I| = 0, \label{eigenvalues}
\end{gather}
where $I$ is an identity matrix of order $n \times n$. The product of the real parts of the eigenvalues for any fixed-point solution $X^{*}$ is
\begin{equation}
    F(X^{*}) = \prod_{i=1}^{n} \Re(\lambda_{i}(X^{*})).
\end{equation}
If the real part of any eigenvalue corresponding to $X^{*}$ is zero, $F(X^{*})$ will be zero. Therefore, the fixed-point solutions satisfying $F(X^{*}) = 0$ correspond to the bifurcation locations.

\subsection{\label{sec:min KL}Minimizing Kullback-Leibler (KL) divergence}
We obtain the bifurcation plot by solving the system of equations (\ref{R_ddot_normalized}), (\ref{phi_ddot_normalized}), and (\ref{rk_dot}) - (\ref{phi_0k equation}) for the fixed-point solutions. These equations require the parameters $c_{k}$, $\omega_{k}$ and $\gamma_{k}$ of the Lorentzian distributions in Eq. \eqref{lorentz_mix}. For accurate results, we learn these parameters by minimizing Kullback-Leibler (KL) divergence between the experimentally obtained and model frequency distributions, using an optimization algorithm. The KL divergence, also known as relative entropy, is commonly used in information theory to quantify the similarity between two density functions. \cite{kullback1997information} Let $\mathbb{E}$ and $\mathbb{M}$ denote the experimental and model frequency distributions, then the KL divergence between them is expressed as
\begin{equation}
    \mathcal{D}(\mathbb{M}||\mathbb{E}) = \sum_{i=1}^{\mathcal{N}}\mathbb{M}_{i} \log \left( \dfrac{\mathbb{M}_{i}}{\mathbb{E}_{i}} \right),
    \label{KL divergence}
\end{equation}
where $\mathbb{E}_{i}$ and $\mathbb{M}_{i}$ are probabilities corresponding to the frequency $f_{i} \in \{f_{1}, f_{2},...,f_{\mathcal{N}}\}$. Here, $f_{1} = 0$ and $f_{\mathcal{N}}$ is the maximum frequency. $\mathbb{E}_{i}$ is obtained for discrete frequencies $f_{i}$ from the FFT of the experimentally obtained heat release rate data. The model frequency distribution depends on the  frequency $f_{i}$ and the vector of parameters $\mathbf{a} = [c_{1}, c_{2}, . . ., c_{n-1}, \gamma_{1}, \gamma_{2}, ... , \gamma_{n}, \omega_{1}, \omega_{2}, ... , \omega_{n}]$ as $\mathbb{M}_{i} = \mathbb{M}(f_{i},\mathbf{a})$, where $\mathbb{M}$ is given by Eq. \eqref{lorentz_mix}. Here, $c_{n}$ is eliminated from $\mathbf{a}$ using Eq. \eqref{fd normalization}. The parameter vector $\mathbf{a}$ is obtained by minimizing the KL divergence $\mathcal{D}$ using the gradient descent method \cite{boyd2004convex}
\begin{equation}
    \mathbf{a}_{i+1} = \mathbf{a}_{i} - \sigma_{L} \nabla_{\mathbf{a}}\mathcal{D},
    \label{gradient descent}
\end{equation}
where $\sigma_{L}$ is the learning rate. We use learning rates of $10^{-6}$ and $10^{-5}$ for frequency distributions of the annular and dump combustor, respectively. The gradient ($\nabla_{\mathbf{a}}\mathcal{D}$) of $\mathcal{D}$ with respect to $\mathbf{a}$ was evaluated using automatic differentiation. \cite{baydin2018automatic}

\section{\label{sec:validation}Validation of the reduced order model}
\begin{figure}
\includegraphics[width=\linewidth]{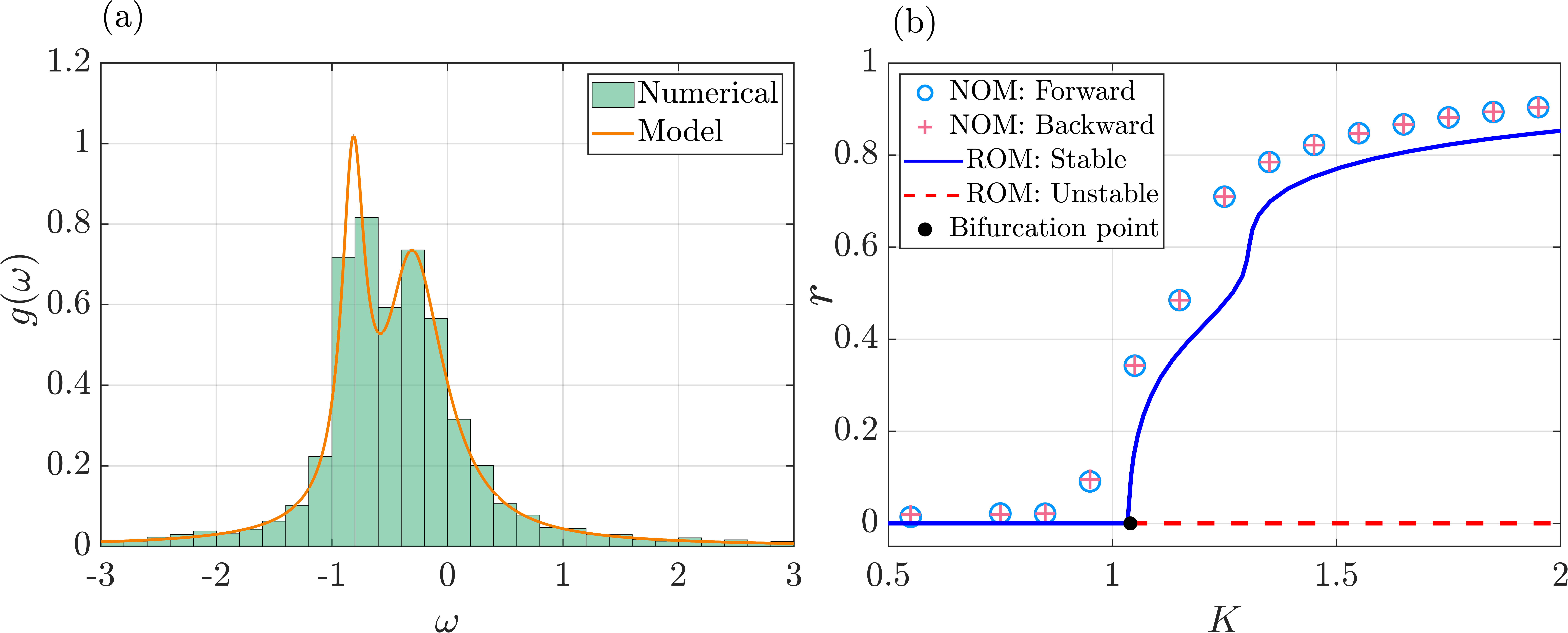}
\caption{\doublespacing (a) Numerical (light green shaded) and model (orange line) natural frequency distributions ($g(\omega)$) of the Kuramoto phase oscillators. The model distribution is given by a sum of three Lorentzian distributions (Eq. \ref{lorentz_mix}). The parameters for each Lorentz distribution are ($\omega_1$, $\omega_2$, $\omega_3$, $\gamma_1$, $\gamma_2$, $\gamma_3$) = (-0.8156, -0.3292, -0.2112, 0.1208, 0.2476, 0.3588) and each Lorentz distribution is weighted by ($c_1$, $c_2$, $c_3$) = (0.3236, 0.2906, 0.3858). (b) Continuous bifurcation of the magnitude of the order parameter ($r$) vs. the coupling strength ($K$). The solid blue and dashed red lines correspond to the stable and unstable fixed-point solutions of the reduced order model (ROM). The light blue circles and the pink plus correspond to the forward and the backward paths, respectively, for the $N$ = 5000 oscillator model (NOM).}
\label{dummy continuous transition}
\end{figure}

\begin{figure}
\includegraphics[width=\columnwidth]{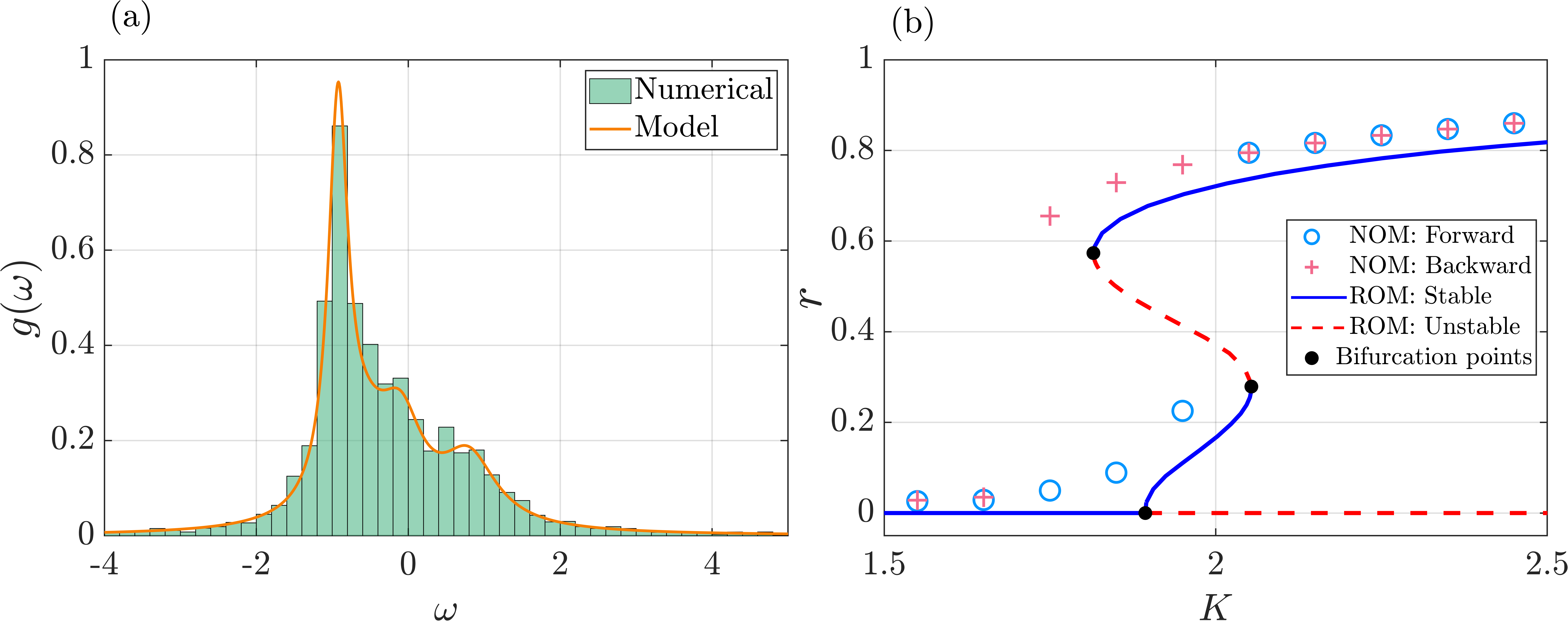}
\caption{\doublespacing (a) Numerical (light green shaded) and model (orange line) natural frequency distribution ($g(\omega)$) of the Kuramoto phase oscillators. The model distribution is given by a sum of three Lorentzian distributions (Eq. \ref{lorentz_mix}). The parameters for each Lorentz distribution are ($\omega_1$, $\omega_2$, $\omega_3$, $\omega_4$, $\gamma_1$, $\gamma_2$, $\gamma_3$, $\gamma_4$) = (-0.9232, -0.0993, 0.7966, -0.5786, 0.1557, 0.3531, 0.4419, 0.4431), and each Lorentz distribution is weighted by ($c_1$, $c_2$, $c_3$, $c_4$) = (0.4063, 0.2056, 0.1986, 0.1896). (b) Secondary bifurcation of the magnitude of the order parameter ($r$) vs. the coupling strength ($K$). The solid blue and dashed red lines correspond to the stable and unstable fixed-point solutions. The light blue circles and the pink plus correspond to the forward and the backward paths, respectively, for the $N$ = 5000 oscillator model (NOM).}
\label{dummy abrupt transition}
\end{figure}

In this section, we validate the reduced order model by comparing the bifurcation plots of fixed point solutions and the numerical simulations of the $N$-oscillator model. The fixed-point solutions are obtained by solving the system of Eqs. (\ref{R_ddot_normalized}) -(\ref{phi_ddot_normalized}) and (\ref{rk_dot}) -(\ref{phi_0k equation}) using the arclength continuation method. We perform the simulations of the $N$-oscillator model by using the fourth-order Runge Kutta (RK4) method on Eqs. (\ref{theta_dot}), (\ref{R_ddot_normalized}) and (\ref{phi_ddot_normalized}) for $N$ = 5000 oscillators and a time step of 0.01. The phases of the oscillators are initialized to be randomly distributed on a circle. After each time step, the magnitude of the order parameter ($r$) is computed using Eq. \eqref{order parameter kuramoto} to obtain a time series of $r$. After removing the initial transience from this time series, we compute the time average and plot this time-averaged value of $r$ with coupling strength (Figs. \ref{dummy continuous transition}(b) and \ref{dummy abrupt transition}(b)). The coupling strength is increased to obtain the forward path and subsequently decreased to obtain the backward path of the bifurcation diagram in steps of 0.1, with the final state taken as the initial state for the next coupling strength. 

Figure \ref{dummy continuous transition}(a) shows the frequency distribution used to obtain the bifurcation plot shown in Fig. \ref{dummy continuous transition}(b). The numerical frequency distribution (light green shaded) and the model frequency distribution (orange line) shown in Fig. \ref{dummy continuous transition}(a) are used in the $N$-oscillator model (NOM) and the reduced order model (ROM), respectively. We increase the coupling strength from 0.55 to 2 for the forward path and decrease it from 2 to 0.55 for the backward path. The system shows a continuous transition from an incoherent state to a synchronized state, which is validated by the absence of hysteresis in forward and backward paths in Fig. \ref{dummy continuous transition}(b).

Figure \ref{dummy abrupt transition}(a) shows the frequency distribution used to obtain the bifurcation plot shown in Fig. \ref{dummy abrupt transition}(b). The numerical frequency distribution (light green shaded) and the model frequency distribution (orange line) shown in Fig. \ref{dummy abrupt transition}(a) are used in the $N$-oscillator model (NOM) and the reduced order model (ROM), respectively. We increase the coupling strength from 1.55 to 2.55 for the forward path and decrease it from 2.55 to 1.55 for the backward path. The system shows a primary bifurcation followed by a secondary bifurcation to high-amplitude limit cycle oscillations. As the coupling strength increases, the system continuously transitions from an incoherent state to a partially synchronized state. On further increasing the coupling strength, the system shows an abrupt transition. 
For stability analysis of the fixed point solutions, we compute the Jacobian matrix and eigenvalues for the system of Eqs. \eqref{R_ddot_normalized}, \eqref{phi_ddot_normalized}, and \eqref{rk_dot} - \eqref{phi_0k equation} using Eqs. \eqref{Jacobian} and \eqref{eigenvalues}. For a fixed point to be stable, the real parts of all its eigenvalues must be negative. If any of the real parts are positive, the fixed point becomes unstable. The stability analysis of the solution branch (indicated by solid blue and dashed red lines in Fig. \ref{dummy abrupt transition}(b)) shows the existence of a bistable region, which is confirmed by the hysteresis in the RK4 simulations.

From Figs. \ref{dummy continuous transition}(b) and \ref{dummy abrupt transition}(b), the bifurcation plots obtained from the reduced order model using the arclength continuation match the bifurcation plots obtained from the RK4 simulations of the $N$ oscillator model for both smooth and abrupt secondary bifurcations. 
Furthermore, the stability of the solution branch, as estimated using the Jacobian, also matches the results from the $N$-oscillator model simulations, thereby validating the reduced-order model.

\section{\label{sec:Results}Results and discussion}
\begin{figure}
    \centering
    \includegraphics[width=\linewidth]{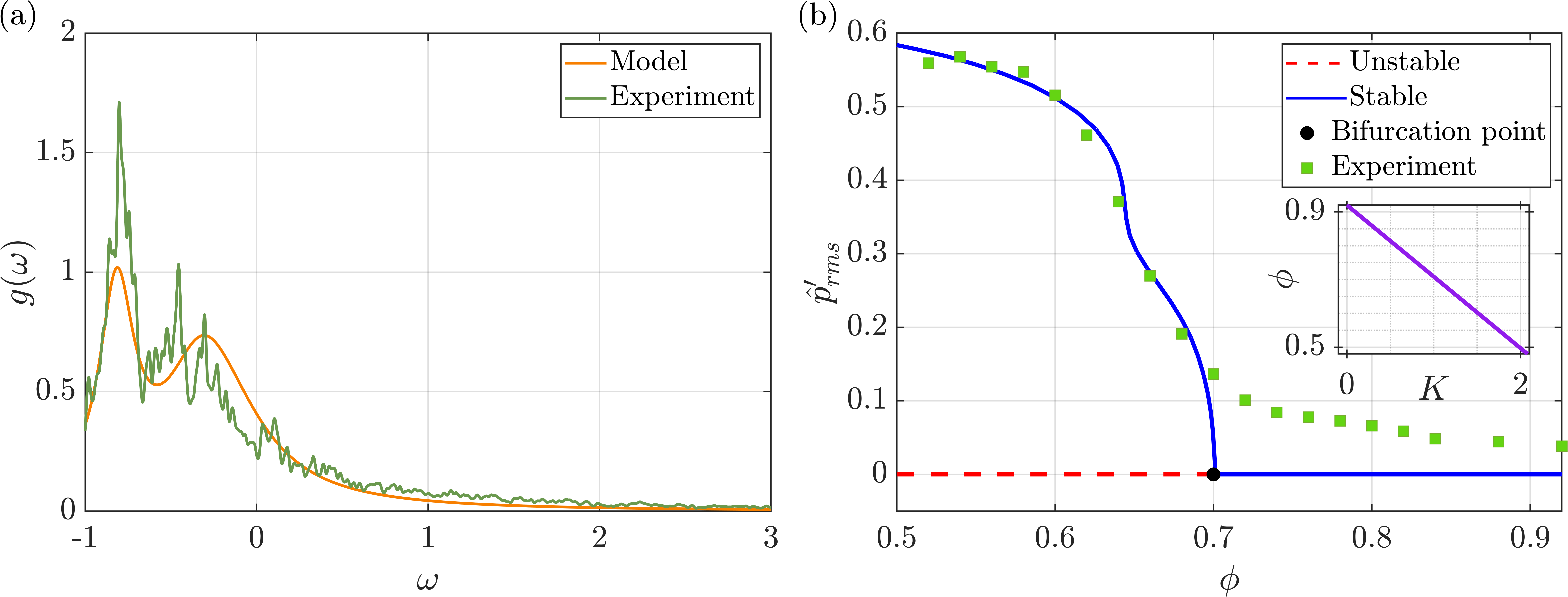}
    \caption{\doublespacing (a) Plot of natural frequency distribution ($g(\omega)$) for phase oscillators. The green line represents the experimentally obtained frequency distribution and the orange line represents the model frequency distribution obtained by using a sum of three Lorentzian distributions. The parameters for each Lorentz distribution are ($\omega_1$, $\omega_2$, $\omega_3$, $\gamma_1$, $\gamma_2$, $\gamma_3$) = (-0.8156, -0.3292, -0.2112, 0.1208, 0.2476, 0.3588) and each Lorentz distribution is weighted by ($c_1$, $c_2$, $c_3$) = (0.3236, 0.2906, 0.3858). (b) Plot of continuous bifurcation of normalized pressure amplitude ($\hat{p}'_{rms}$) with equivalence ratio ($\phi$). Solid blue and dashed red lines represent stable and unstable fixed-point solutions, respectively. Green squares represent experimental data for the forward path, obtained from the bluff body stabilized dump combustor. Inset shows the linear relation between the coupling strength ($K$) and $\phi$, given as $\phi = -0.2115K + 0.92$.}
    \label{fig:continuous transition}
\end{figure}

\begin{figure}
    \centering
    \includegraphics[width=\linewidth]{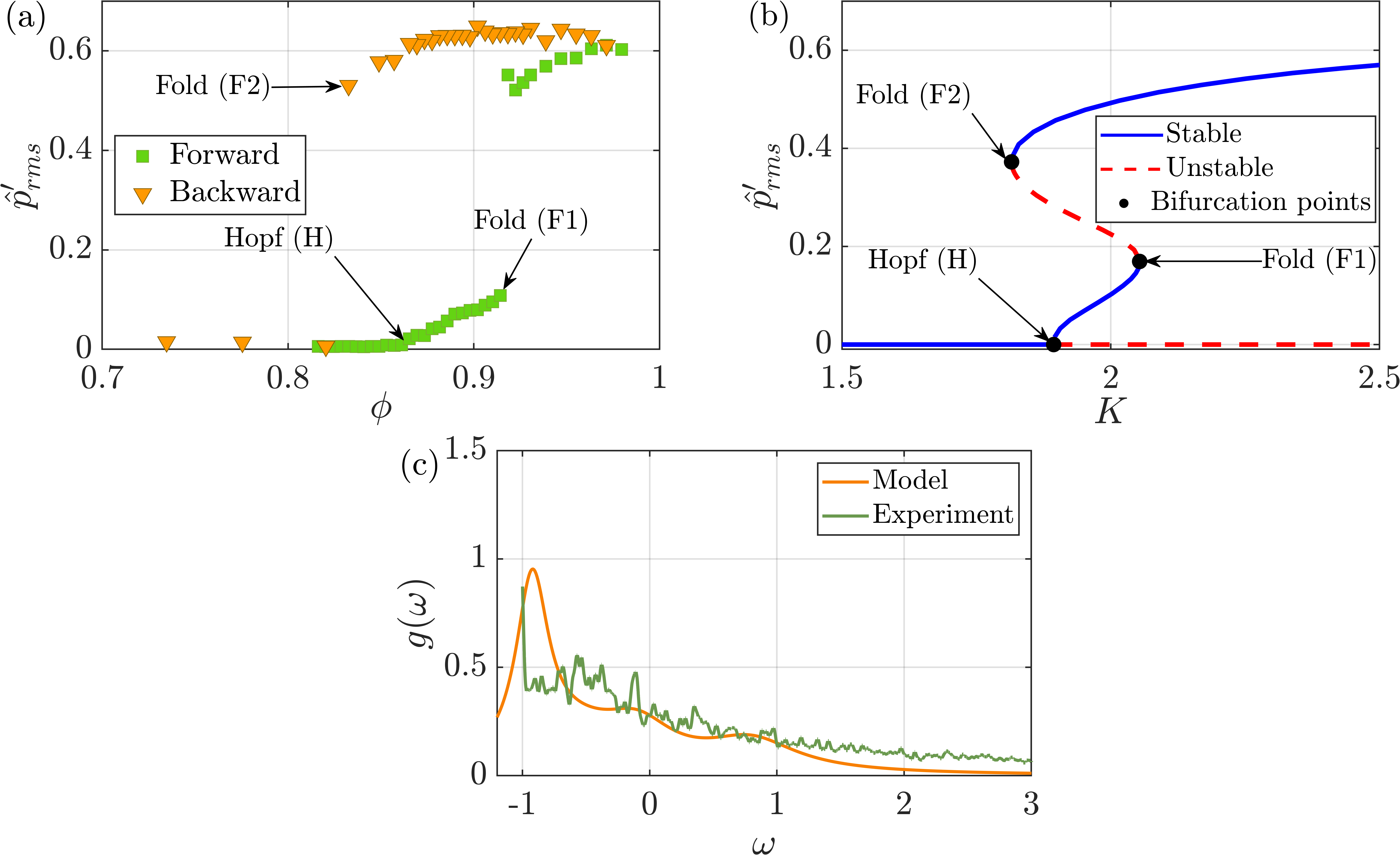}
    \caption{ \doublespacing (a) Secondary bifurcation of normalized pressure amplitude ($\hat{p}'_{rms}$) with equivalence ratio ($\phi$) obtained from swirl-stabilized annular combustor. Green squares and orange triangles represent experimental data for forward and backward paths, respectively, obtained from the annular combustor. (b) secondary bifurcation of $\hat{p}'_{rms}$ with coupling strength ($K$) obtained from the model. Solid blue and dashed red lines represent stable and unstable fixed-point solutions, respectively. (c) Plot of natural frequency distribution ($g(\omega)$) for phase oscillators. The green line represents the experimentally obtained frequency distribution, and the orange line represents the model frequency distribution obtained by using a sum of four Lorentzian distributions. The parameters for each Lorentz distribution are ($\omega_1$, $\omega_2$, $\omega_3$, $\omega_4$, $\gamma_1$, $\gamma_2$, $\gamma_3$, $\gamma_4$) = (-0.9232, -0.0993, 0.7966, -0.5786, 0.1557, 0.3531, 0.4419, 0.4431), and each Lorentz distribution is weighted by ($c_1$, $c_2$, $c_3$, $c_4$) = (0.4063, 0.2056, 0.1986, 0.1896). }
    \label{fig:secondary transition}
\end{figure}

We start by comparing the model predictions with the experimental results for two combustors, viz., bluff-body stabilized dump combustor and swirl-stabilized annular combustor. Figures \ref{fig:continuous transition}(a) and \ref{fig:secondary transition}(c) (green lines) show the frequency distribution ($g(\omega)$) obtained from the FFT of the heat release rate data obtained from experiments during combustion noise ($\phi = 0.82$ for swirl-stabilized annular combustor and $\phi = 0.92$ for bluff body stabilized dump combustor). These frequency distributions obtained from experiments are approximated using model frequency distributions (orange lines in Figs. \ref{fig:continuous transition}(a) and \ref{fig:secondary transition}(c)) given by Eq. \eqref{lorentz_mix} with $n$ = 3 and $n$ = 4 for dump combustor and annular combustor, respectively. The parameters ($c_{k}, \omega_{k}$ and $\gamma_{k}$) of the model distribution are obtained by minimizing the KL divergence. Using these parameters in Eqs. (\ref{R_ddot_normalized})-(\ref{phi_ddot_normalized}) and (\ref{rk_dot})-(\ref{phi_0k equation}), we compute and plot the fixed-point solutions. The method to obtain the normalized pressure amplitude ($\hat{p}'_{rms}$) for the reduced model at the fixed point solutions is described in Appendix \ref{appendix: p_rms derivation}. 
% The bluff-body stabilized dump combustor exhibits a continuous bifurcation (Fig. \ref{fig:continuous transition}) when the equivalence ratio is decreased. On the other hand, the annular combustor exhibits a primary bifurcation followed by a secondary bifurcation to high-amplitude thermoacoustic instability when the equivalence ratio is increased, as shown in Fig. \ref{fig:secondary transition}. 

Figures \ref{fig:continuous transition}(a) and \ref{fig:continuous transition}(b) show the natural frequency distribution for the oscillators and the corresponding bifurcation plot for the bluff body stabilized dump combustor. We assume a linear relation between the coupling strength ($K$) and the equivalence ratio ($\phi$) of the form $\phi = mK +c$. The parameters $m$ and $c$ are obtained by matching $K = 0$ to $\phi = 0.92$ which corresponds to the dynamical state of combustion noise and $K = 0.998$ to $\phi = 0.70$ which corresponds to the bifurcation point. In Fig. \ref{fig:continuous transition}(b), the system transitions continuously from combustion noise to thermoacoustic instability when the equivalence ratio is decreased. The stable fixed-point solutions after the bifurcation match the experimentally obtained bifurcation diagram, which confirms the linear relation between $K$ and $\phi$.

Figure \ref{fig:secondary transition}(a) shows the bifurcation of normalized pressure amplitude with equivalence ratio obtained from experiments. When $\phi$ is increased, the system transitions continuously from combustion noise to low-amplitude limit cycle oscillations, followed by an abrupt jump to high-amplitude limit cycle oscillations. Figure \ref{fig:secondary transition}(b) shows the bifurcation plot of normalized pressure amplitude with the coupling strength, obtained from the model using the frequency distribution shown in Fig. \ref{fig:secondary transition}(c) (orange line). The model predicts the secondary transition observed in the experiments. However, unlike the continuous case, we do not observe a linear relation between the normalized equivalence ratio and the coupling strength. Hence, to verify the quality of the prediction, we define a ratio $\xi$ as
\begin{equation}
    \xi = \dfrac{x_{F1} - x_{F2}}{x_{F1} - x_{H}},
\end{equation}
where $x$ is the equivalence ratio ($\phi$) for the experimental case and the coupling strength ($K$) for the model. $x_{H}$, $x_{F1}$ and $x_{F2}$ correspond to the Hopf point (H), fold point (F1) and fold point (F2) of the bifurcations, respectively, as shown in Figs. \ref{fig:secondary transition}(a) and \ref{fig:secondary transition}(b). The ratio $\xi$ calculated using the experimental and model data is $\xi_{expt} = 1.5385$ and $\xi_{model} = 1.4893$. 
The values of $\xi$ for the model and the experiment are close, which shows that it is possible to predict the location of the backward jump ($\phi_{F2}$) using $\xi_{model}$ if the locations of Hopf point ($\phi_{H}$) and forward jump ($\phi_{F1}$) are known.
The model is able to capture the jump locations qualitatively. To check whether the model captures the jump in amplitudes, we compare the normalized pressure amplitude ($\hat{p}'_{rms}$) just after the jump, which comes out to be 0.521 and 0.505 for the experimental and model bifurcations, respectively. Since both $\xi$ and the jump amplitude closely match between the model and the experiment, we conclude that the model has qualitatively captured the experimental bifurcation.

\section{\label{sec:conclusion}Conclusion}
We presented a reduced-order mean-field model to capture the transitions to thermoacoustic instability with a minimum number of equations. The reduced order model captures the experimentally observed continuous and abrupt secondary transitions in the bluff-body stabilized dump combustor and the swirl-stabilized annular combustor. 

We modeled the heat release rate as an ensemble of coupled Kuramoto phase oscillators. Using the Ott-Antonsen ansatz, we obtain a reduced-order model, which is then validated by comparing the bifurcations with the $N$-oscillator model. The natural frequency distribution for the oscillators is obtained from the FFT of the heat release rate fluctuations during the occurrence of combustion noise. This natural frequency distribution obtained from experiments is approximated using a sum of Lorentzian distributions by minimizing KL divergence. We observed a linear relation between the equivalence ratio and the coupling strength for the bluff-body stabilized dump combustor. Such a linear relation is not followed in the swirl-stabilized annular combustor. A qualitative comparison between the experimental and model bifurcation diagrams, using parameters such as jump amplitudes and the hysteresis width ratios, showed excellent agreement, confirming that the model captures the experimental bifurcations accurately. The relation between the equivalence ratio and coupling strength requires further investigation, as it may depend on the combustor geometry or burner configuration.

The capability of the reduced order model to correctly capture the different bifurcations holds the potential to study the effects of different frequency distributions on the nature of bifurcation and develop control strategies for practical thermoacoustic systems.
As demonstrated, the method to approximate the natural frequency distribution observed in experiments using Lorentzian distributions is effective and can make the Ott-Antonsen reduction method applicable to other dynamical systems containing Kuramoto oscillators whose natural frequencies are obtained from experiments.

\begin{acknowledgments}
We thank Ramesh S. Bhavi and S. Sudarsanan for providing the experimental data. We acknowledge B. Thonti and S. Sudarsanan for their valuable comments. S. Singh and J. Dhadphale gratefully acknowledge the Ministry of Human Resource Development (MHRD) for Ph.D. funding through the Half-Time Research Assistantship (HTRA). J. Dhadphale is grateful to the Ministry of Education for providing the fellowship under the Prime Minister’s Research Fellows (PMRF) scheme. R. I. Sujith acknowledges the funding from the IOE initiative (No. SP22231222CPETWOCTSHOC).
\end{acknowledgments}

\section*{Author Declarations}
\subsection*{Conflict of Interest}
The authors have no conflict to disclose.

\section*{Data Availability Statement}
The data that support the findings of this study are available
from the corresponding authors upon reasonable request.

\appendix

\section{\label{appendix: derivation of Rddot and phiddot}Derivation of second order differential equations for $R$ and $\Phi$}
Equations \eqref{R_ddot_nonnormalized} and \eqref{phi_ddot_nonnormalized} are obtained by substituting Eq. \eqref{Krylov Bogolyubov decomposition} in Eq. \eqref{thermoacoustic}. Using Eq. \eqref{Krylov Bogolyubov decomposition} we obtain the time derivatives of $\eta$,
\begin{gather}
    \dot{\eta} = -\dot{R} \cos(t + \Phi) + R (1 + \dot{\Phi}) \sin(t+\Phi),  \label{eta_dot}\\
    \ddot{\eta} = -\ddot{R} \cos(t+\Phi) + R (1+\dot{\Phi})^2 \cos(t+\Phi) +  2\dot{R} (1+\dot{\Phi}) \sin(t+\Phi) +R \ddot{\Phi} \sin(t+\Phi) .
\end{gather}
Substituting $\dot{\eta}$ and $\ddot{\eta}$ from the above equations in Eq. \eqref{thermoacoustic} and rearranging the terms, we get the left-hand side (LHS) and right-hand side (RHS) of the resulting equations as
\begin{gather}
    \text{LHS} = [-\ddot{R} + R(1+\dot{\Phi})^2 -\alpha \dot{R} - R]\cos(t+\Phi) + [2\dot{R}(1+\dot{\Phi}) + R\ddot{\Phi} + \alpha R (1+\dot{\Phi})]\sin(t+\Phi), \label{LHS}\\
    \text{RHS} = \beta \cos(kz_{f})\sum_{j=1}^{N}\sin(t+\theta_{j}). \label{RHS}
\end{gather}
For convenience, we re-write the coefficients of sines and cosines in Eq. \eqref{LHS} as
\begin{gather}
    \mathcal{A} = -\ddot{R} + R(1+\dot{\Phi})^2 -\alpha \dot{R} - R, \label{cos coeff}\\
    \mathcal{B} = 2\dot{R}(1+\dot{\Phi}) + R\ddot{\Phi} + \alpha R (1+\dot{\Phi}), \label{sin coeff}
\end{gather}
which leads to 
\begin{gather}
    \text{LHS} = \mathcal{A} \cos{(t+\Phi)} + \mathcal{B} \sin{(t+\Phi)}. \label{LHS modified}
\end{gather}
For an arbitrary angle $\Psi$, sines and cosines can be expressed in complex form  using Euler's representation as
\begin{gather}
    \sin(\Psi) = \dfrac{e^{i\Psi} - e^{-i\Psi}}{2i}, \\
    \cos(\Psi) = \dfrac{e^{i\Psi} + e^{-i\Psi}}{2}.
\end{gather}
Using this complex representation for sines and cosines in Eqs. \eqref{RHS} and \eqref{LHS modified} leads to
\begin{gather}
    \mathcal{A} \left[ \dfrac{e^{i(t+\Phi)} + e^{-i(t+\Phi)}}{2} \right] + \mathcal{B} \left[ \dfrac{e^{i(t+\Phi)} - e^{-i(t+\Phi)}}{2i} \right] = \beta \cos(kz_{f})\sum_{j=1}^{N}\left[ \dfrac{e^{i(t+\theta_{j})} - e^{-i(t+\theta_{j})}}{2i} \right].
\end{gather}
Multiplying both sides by $e^{-it}$ leads to
\begin{gather}
    \mathcal{A} \left[ \dfrac{e^{i\Phi} + e^{-2it}e^{i\Phi}}{2} \right] + \mathcal{B} \left[ \dfrac{e^{i\Phi} - e^{-2it}e^{\Phi}}{2i} \right] = \beta \cos(kz_{f})\sum_{j=1}^{N}\left[ \dfrac{e^{i\theta_{j}} - e^{-2it}e^{\theta_{j}}}{2i} \right],
\end{gather}
computing the time average over a period of $2\pi$, assuming the variation of $R$ and $\Phi$ to be slow over the period of $T=2\pi$, 
\begin{equation}
    \mathcal{A} \dfrac{e^{i\Phi}}{2} + \mathcal{B} \dfrac{e^{i\Phi}}{2i} = \beta \cos(kz_{f})\sum_{j=1}^{N} \dfrac{e^{i\theta_j}}{2i}.
\end{equation}
Re-substituting $\mathcal{A}$ and $\mathcal{B}$ from Eqs. \eqref{cos coeff} and \eqref{sin coeff} and equating the imaginary and real parts yields the second-order ordinary differential equations for $R$ and $\Phi$ expressed by Eqs. \eqref{R_ddot_nonnormalized} and \eqref{phi_ddot_nonnormalized}.

\section{\label{appendix: p_rms derivation} Extraction of normalized pressure amplitude ($\hat{p}'_{rms}$) from $R$ and $\Phi$}
The Galerkin modal expansion of acoustic pressure fluctuations is given by Eq. \eqref{pprime}. As discussed in Sec. \ref{sec:method}, we neglect the contribution from higher modes and non-dimensionalize Eq. \eqref{pprime} using Eq. \eqref{nondimensionalization}
\begin{equation}
    p' = \dot{\eta} \, \cos(kz),
\end{equation}
where $p' = \tilde{p}'/\tilde{p}_{0}$. The normalized pressure fluctuations are given by
\begin{gather}
    \hat{p}' = \dfrac{p'}{p'_{LCO,amp}}= \dfrac{\dot{\eta}}{\dot{\eta}_{LCO,amp}}, \label{phatprime}
\end{gather}
where $\dot{\eta}_{LCO,amp}$ and $p'_{LCO,amp}$ are the amplitudes of $\dot{\eta}$ and $p'$ at limit cycle oscillations for $K \to \infty$. From Eq. \eqref{eta_dot}, $\dot{\eta}_{LCO}$ is
\begin{equation}
    \dot{\eta}_{LCO} = -\dot{R}_{LCO} \, \cos(t+\Phi_{LCO}) + R_{LCO}(1+\dot{\Phi}_{LCO}) \sin(t+\Phi_{LCO}). \label{eta_dot_appendix}
\end{equation}
For the limit cycle and synchronization conditions discussed in Sec. \ref{sec:method}, we can show that $\dot{\Phi}_{LCO} = 0$ from Eq. \eqref{phi_ddot_normalized} and $\dot{R}_{LCO} = 0$ for the fixed point condition. This gives 
\begin{gather}
\dot{\eta}_{LCO} = R_{LCO} \sin(t + \Phi_{LCO}), \\
\dot{\eta}_{LCO,amp} = R_{LCO}.\label{lco eta}
\end{gather}
From Eqs. \eqref{phatprime} and \eqref{lco eta},
\begin{equation}
    \hat{p}' = \dot{\hat{\eta}}, \label{phatprime_norm}
\end{equation}
where $\hat{\eta} = \eta / R_{LCO}$. From Eq. \eqref{eta_dot}, $\dot{\hat{\eta}}$ is
\begin{gather}
    \dot{\hat{\eta}} = -\dot{\hat{R}} \, \cos(t+\Phi) + \hat{R}(1+\dot{\Phi}) \sin(t+\Phi) \label{eta_hat_dot}.
\end{gather}
At fixed point, $\dot{\hat{R}} = 0$. Substituting Eq. \eqref{eta_hat_dot} in Eq. \eqref{phatprime_norm} and computing the root mean square (rms), we obtain the normalized pressure amplitude ($\hat{p}'_{rms}$) at fixed points as
\begin{equation}
    \hat{p}'_{rms} = \dfrac{\hat{R}(1+\dot{\Phi})}{\sqrt{2}}.
\end{equation}

%%%%%%%%%%%%%%%%%%%%%%%%%%%%%%%%%%%%%%%
\bibliography{manuscript}% Produces the bibliography via BibTeX.

\end{document}